\documentclass[twocolumn,aps,superscriptaddress,floatfix,bibnotes,nobibnotes]{revtex4-1}

\usepackage[colorlinks=true,citecolor=blue,urlcolor=blue,linkcolor=black]{hyperref}
\usepackage[version=3]{mhchem}
\usepackage{graphicx}

\usepackage[dvipsnames]{xcolor}

\begin{document}
	\title{Nematic response revealed by coherent phonon oscillations in BaFe\textsubscript{2}As\textsubscript{2}}
		
	\author{Min-Cheol Lee}
    \thanks{Corresponding author}
    \email{mclee@lanl.gov}
	\affiliation{Center for Correlated Electron Systems (CCES), Institute for Basic Science (IBS), Seoul 08826, Republic of Korea}
	\affiliation{Department of Physics and Astronomy, Seoul National University, Seoul 08826, Republic of Korea}
	\affiliation{Center for Integrated Nanotechnologies, Los Alamos National Laboratory, Los Alamos, New Mexico 87545, USA}
	
	\author{Inho Kwak}
	\affiliation{Center for Correlated Electron Systems (CCES), Institute for Basic Science (IBS), Seoul 08826, Republic of Korea}
	\affiliation{Department of Physics and Astronomy, Seoul National University, Seoul 08826, Republic of Korea}
	\author{Yeongseon Lee}
	\affiliation{Department of Physics, Chungbuk National University, Cheongju, Chungbuk 28644, Republic of Korea}
	\author{Bumjoo Lee}
	\author{Byung Cheol Park}
	\affiliation{Center for Correlated Electron Systems (CCES), Institute for Basic Science (IBS), Seoul 08826, Republic of Korea}
	\affiliation{Department of Physics and Astronomy, Seoul National University, Seoul 08826, Republic of Korea}
	\author{Thomas Wolf}
	\affiliation{Institute for Quantum Materials and Technologies, Karlsruhe Institute of Technology, 76021 Karlsruhe, Germany}
	\author{Tae Won Noh}
	\affiliation{Center for Correlated Electron Systems (CCES), Institute for Basic Science (IBS), Seoul 08826, Republic of Korea}
	\affiliation{Department of Physics and Astronomy, Seoul National University, Seoul 08826, Republic of Korea}
	\author{Kyungwan Kim}
	\thanks{Corresponding author}
	\email{kyungwan@chungbuk.ac.kr}
	\affiliation{Department of Physics, Chungbuk National University, Cheongju, Chungbuk 28644, Republic of Korea}

	\date{\today}

\begin{abstract}
We investigate coherent phonon oscillations of BaFe\textsubscript{2}As\textsubscript{2} using optical pump-probe spectroscopy. Time-resolved optical reflectivity shows periodic modulations due to $A_{1g}$ coherent phonon of $c$-axis arsenic vibrations. Optical probe beams polarized along the orthorhombic $a$- and $b$-axes reveal that the initial phase of coherent oscillations shows a systematic deviation as a function of temperature, although these oscillations arise from the same $c$-axis arsenic vibrations. The oscillation-phase remains anisotropic even in the tetragonal structure, reflecting a nematic response of BaFe\textsubscript{2}As\textsubscript{2}. Our study suggests that investigation on the phase of coherent phonon oscillations in optical reflectivity can offer unique evidence of a nematic order strongly coupled to a lattice instability.
\end{abstract}
\maketitle
 One of recent interests in the phase diagram of pnictide superconductors is a nematic state below $T_\textrm{nem}$, which is represented by a broken \textit{C}\textsubscript{4} symmetry even with spin-rotational symmetry \cite{Paglione2010,Johnston2010,Fernandes2014,Fernandes2012}. The nematic order becomes static together with the phase transitions of tetragonal-to-orthorhombic structural distortion at $T_\textrm{S}$, antiferromagnetic spin ordering at $T_\textrm{N}$, and superconducting ordering at $T_\textrm{C}$; $T_\textrm{nem}=T_\textrm{S}> T_\textrm{N}>T_\textrm{C}$. However, the nematic fluctuations leave its footprints in various physical quantities, such as dc resistivity \cite{Chu2008}, magnetic torque \cite{Kasahara2012}, electronic and orbital structures \cite{Yi2011,Kim2013}, and optical conductivity \cite{Nakajima2011,Mirri2015} with a strong \textit{C}\textsubscript{2} anisotropy between \textit{x} and \textit{y} directions of the Fe-plane. Theoretical predictions have suggested that the nematicity arises from short-range spin fluctutaions as well as orbital fluctuations with discrepancy in $d$-orbital occupancies \cite{Fernandes2012,Fernandes2014}. However, the relation between the nematic fluctuations and structural instability has not been clarified.
	
 Coherent phonon oscillations, in-phase lattice vibrations triggered by an ultrashort optical pulse, have provided a unique perspective on crystal structures \cite{Zeiger1992,Kuznetsov1994,Stevens2002,Riffe2007,Lee2019_1,Lee2019_2}. BaFe\textsubscript{2}As\textsubscript{2}, a parent compound of pnictide superconductors, presents robust features of a coherent phonon with a frequency of 5.45 THz that is consistent with an $A_{1g}$ phonon mode of arsenic vibrations along $c$-axis (Fig. 1(a)) \cite{Gerber2015,Rettig2015,Mansart2010,Kim2012,Yang2014,Okazaki2018}. More specifically, coherent oscillations of the $A_{1g}$ mode have been observed in various probing signals such as X-ray diffraction intensity \cite{Rettig2015,Gerber2015} and photoemission spectra \cite{Yang2014,Okazaki2018} as well as optical properties \cite{Mansart2010,Kim2012}, indicating a strong coupling of the $A_{1g}$ phonon to the electronic structures. This $A_{1g}$ phonon is also coupled to the magnetic ground states such that the coherent $A_{1g}$ lattice motions induce a transient spin-density-wave state even in the paramagnetic phase above $T_\textrm{N}$ \cite{Kim2012}. Despite these intensive studies on the coherent $A_{1g}$ phonon, it has not been revealed how this phonon couples to the nematic ordering in BaFe\textsubscript{2}As\textsubscript{2} and other pnictide compounds.
	
	\begin{figure}[!b]
	\includegraphics[width=3.4 in]{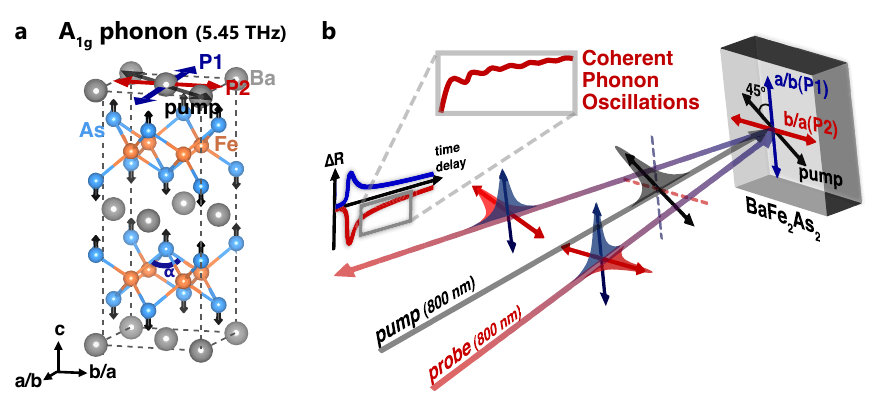}
	\centering
	\caption{(a) The crystal structure of BaFe\textsubscript{2}As\textsubscript{2} and (b) a schematic diagram of optical pump - optical probe spectroscopy. The eigenmode of the $A_{1g}$ phonon with $c$-axis movements of As ions are depicted in (a) indicated as black arrows.}
	\label{FIG1}
	\end{figure}
	
 In this paper, we present an anisotropic response in time-resolved reflectivity due to the $A_{1g}$ phonon oscillations of BaFe\textsubscript{2}As\textsubscript{2}. We found that the initial phase of coherent $A_{1g}$ phonon oscillations exhibits an anisotropic behavior depending on probe light polarization along the orthorhombic axes. Notably, the phase-difference anomaly was clearly observed even in the tetragonal state above $T_\textrm{S}$, suggesting the optical anisotropy in the coherent phonon oscillations is a unique signature of the nematic state in BaFe\textsubscript{2}As\textsubscript{2}.
	
 We employed femtosecond optical pulses to investigate time-resolved reflectivity on BaFe\textsubscript{2}As\textsubscript{2} single crystals. We utilized near infrared 800-nm pulses for both the pump and probe beams generated from a commercial Ti:Sapphire amplifier system with a 250-kHz repetition rate. The duration of the pump and probe pulses were 30 fs. The full width at half maximum spot sizes of the pump and probe pulses were 160 and 60 $\mu$m, respectively. We used pump beams with various absorbed fluences from 24 $\mu$J cm\textsuperscript{-2} to 3 mJ cm\textsuperscript{-2}. The polarizations of the two probe lights (P1 and P2) are along orthorhombic axes of BaFe\textsubscript{2}As\textsubscript{2}, which are parallel to bonding directions ($x$ and $y$) between nearest neighboring Fe ions (Fig. 1(a) and 1(b)). We set the polarization of the pump beam in-between the probe polarizations.
	
\begin{figure}[t]
	\includegraphics[width = 3.4 in]{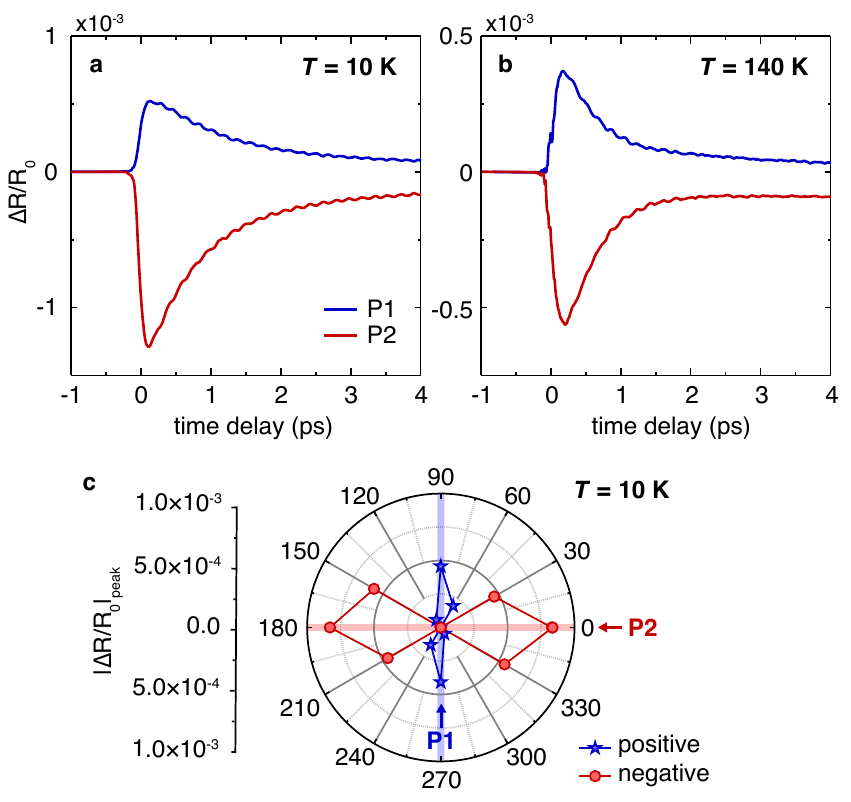}
	\centering
\caption{Near-infrared reflectivity transient at (a) 10 K ($<$ \textit{T}\textsubscript{N} = 134 K) and (b) 140 K ($>$ \textit{T}\textsubscript{N}). The reflectivity transients are obtained with optical probes with different light polarizations along orthorhombic axes with a positive signal (P1; blue) and a negative signal (P2; red) at $T$ = 10 K. (c) Positive maximum (blue) and negative minimum (red) values of $\Delta{R}/R_0$ depending on the probe polarization direction; P1 and P2 are parallel to the in-plane axes of the orthorhombic unit cell, showing positive maximum and negative minimum signals, respectively.}
\label{FIG2}
\end{figure}

 Figure 2 shows time-resolved reflectivity changes of BaFe\textsubscript{2}As\textsubscript{2} triggered by near-infrared pumping at 10 K and 140 K at a fluence of 24 $\mu$J cm\textsuperscript{-2}. The overall dynamics shows clear anisotropy depending on the probe beam polarizations. The linear polarziation of the probe light is indicated by P1 (blue) and P2 (red) of which measurements present transient reflectivity peak values of a positive maximum and a negative minimum, respectively. We determined the P1 and P2 directions through a $d$-wave-like azimuthal dependence of the peak value as shown in Fig. 2(c). We also measured X-ray diffraction at room temperature to confirm that these P1 and P2 directions are along the Fe-Fe bonding directions. These results agree with the previous measurements of BaFe\textsubscript{2}As\textsubscript{2} \cite{Stojchevska2012}, and other pnicitides such as FeSe \cite{Luo2017} and Na(Fe,Co)As \cite{Liu2018}. Such anisotropic responses in transient reflectivity have been suggested as evidence of the nematic order because they appear even in the paramagnetic tetragonal phases \cite{Stojchevska2012,Luo2017,Liu2018}. Our data at 140 K and 160 K (not shown) clearly show the nematic responses with clear anisotropy above $T_\textrm{N}=$ 134 K and $T_\textrm{S}=$ 135 K, consistent with a previous report using the same pnictide compound \cite{Stojchevska2012}.

 \begin{figure}[]
	\includegraphics[width=3.4 in]{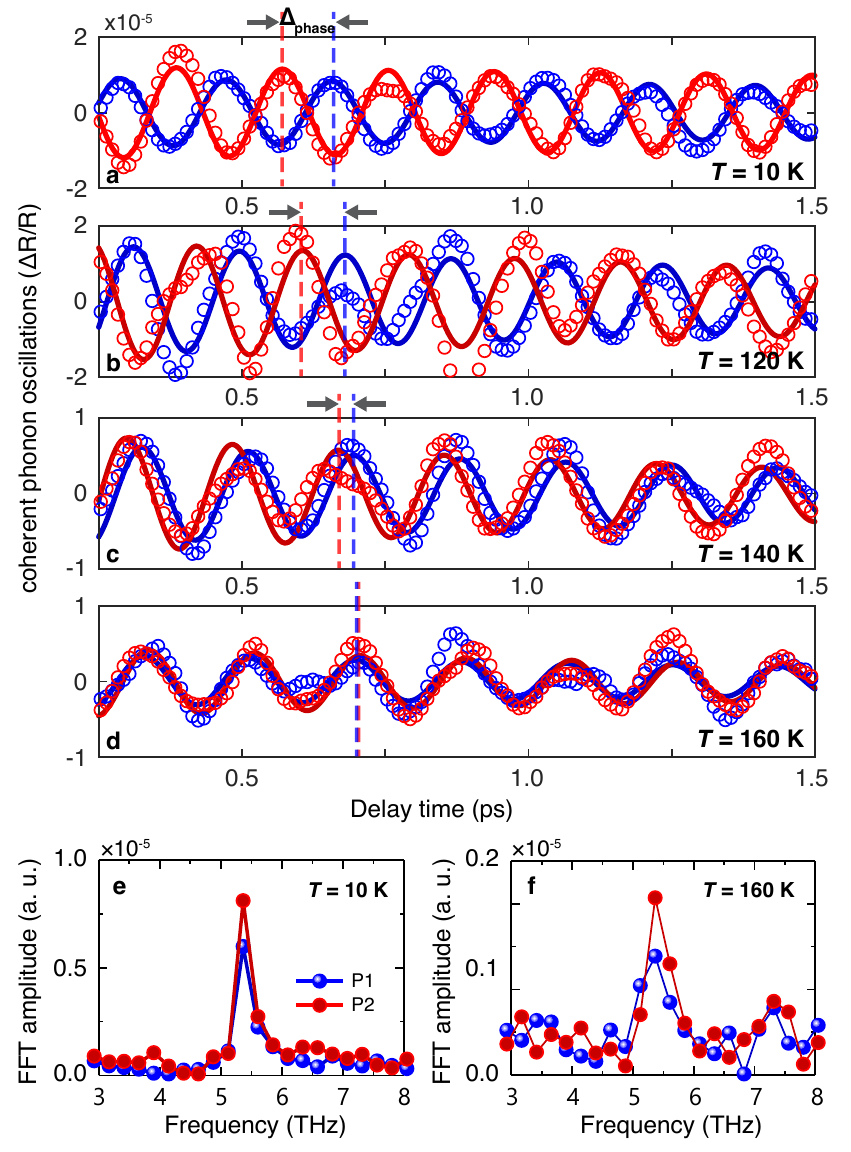}
	\centering
	\caption{Raw data of coherent oscillations of the $A_{1g}$ phonon (open circles) and fit curves of a damped harmonic oscillator model (solid lines) at various temperatures; (a) 10 K, (b) 120 K, (c) 140 K, and (d) 160 K. The oscillations are obtained by using probe lights with orthogonal P1 and P2 polarizations as indicated in Fig. 2. Fourier transform of the oscillatory signals at (e) 10 K and (f) 160 K.}
	\label{FIG3}
\end{figure}

 In addition to the overall dynamics, we observed clear periodic oscillations in the transient reflectivity. We extract oscillating components, as shown in Figs. 3(a-d), by subtracting the electronic responses by means of bi-exponential curve fitting. Fourier transform analysis reveals that the coherent oscillations are composed of a single phonon mode of 5.45 THz (Figs. 3(e-f)), which matches the resonant frequency of the $A_{1g}$ Raman active mode displayed in Fig. 1(b) \cite{Rahlenbeck2009, Litvinchuk2008}. Particularly, we find that the phases of the $A_{1g}$ oscillations in P1 and P2 probing signals become distinct at low temperatures. The high-temperature in-phase oscillations gradually turn into out-of-phase ones as temperature decreases. As a result, complete out-of-phase oscillations between the P1 and P2 signals were observed at 10 K.

 For a further quantitative analysis on the oscillation-phase, we fit the data with a damped harmonic oscillator model: $\Delta{R}_{CP}(t) = -{A}\cos(2\pi{f}t-\theta)\exp(-t/\tau)$, where ${A}$, ${f}$, ${\theta}$, and ${\tau}$ present the amplitude, frequency, initial phase, and damping time of the $A_{1g}$ phonon mode, respectively. The fitting results are plotted in Fig. 3(a-d) as solid line curves which all agree well with the measured data.
 
\begin{figure}[]
	\includegraphics[width=3.4 in]{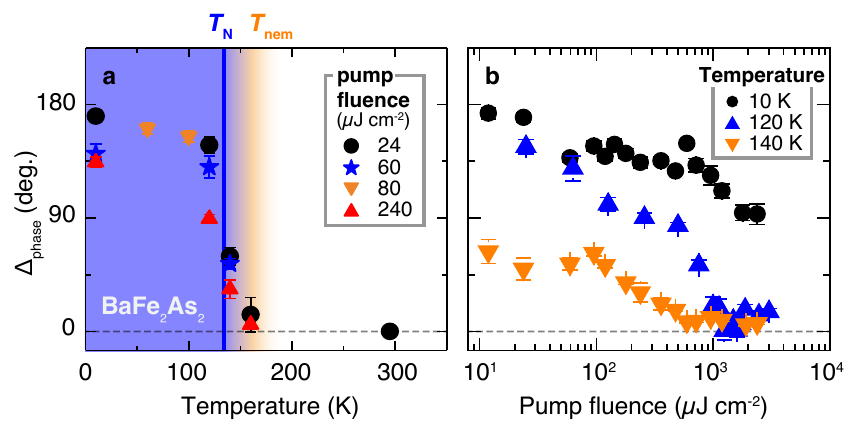}
	\centering
	\caption{Difference of oscillation-phase of the $A_{1g}$ coherent phonon, $\Delta_\textrm{phase} = \theta_\textrm{P1} - \theta_\textrm{P2}$, as a function of (a) temperature and (b) pump fluence. The phase difference presents a finite value in the tetragonal phase at 140 K ($> T_\textrm{S}$ = 135 K).}
	\label{FIG4}
\end{figure}

 Figure 4 shows the difference in the oscillation-phase between P1 and P2 probing signals ($\Delta_{phase} = |\theta_{\textrm{P1}} - \theta_{\textrm{P2}}|$). We focus on the difference value rather than the individual phase values to discuss the anisotropic response. The phase-difference is not fully suppressed at $T$ = 140 K ($> T_\textrm{N}$, $T_\textrm{S}$), where the material stays in the paramagnetic tetragonal state (Fig. 4(a)). The data obtained by various pump fluences indicate that the phase-difference $\Delta_{phase}$ at 140 K presents finite values at low fluences, which decreases under higher fluence pumping as shown Fig. 4(b), suggesting that the nematicity is suppressed under extremely high fluence. We note that the pump-induced heating does not play an important role here because the overall fluence dependence deviates from the temperature dependence. Understanding the details is beyond the scope of this paper and will be discussed in a separated manuscript \cite{Lee2021}. We also find that such finite $\Delta_{phase}$ is also likely to remain at even higher temperature $T = 160$ K, although it is not conclusive because of its large error bar comparable to $\Delta_{phase}$. This oscillation phase of the coherent $A_{1g}$ phonon reminds us of the nematic order.
 
 What is the microscopic origin of the oscillation phase difference? Periodic modulations in reflectivity due to coherent phonon oscillations should follow $\Delta{R}_\textnormal{CP} = (\partial{R}/\partial{Q})\delta{Q}$ \cite{Stevens2002,Riffe2007,Lee2019_1}. More specifically, coherent oscillations in probing signals depend on a displacement in the lattice coordinates ($\delta{Q}$) and a consequent response of reflectivity ($\partial{R}/\partial{Q}$). The initial lattice displacement $\delta{Q}$ is determined by the Raman tensor of a phonon mode, resulting in a driving force upon pumping \cite{Stevens2002,Riffe2007}. In BaFe\textsubscript{2}As\textsubscript{2}, the coherent $A_{1g}$ oscillations in the transient reflectivity ($\Delta{R}_\textnormal{CP}$) do not change depending on the pump polarization direction (not shown). We used the same pump polarization to obtain the data of Fig. 2 and Fig. 3, and therefore the anisotropic responses should not originate from the lattice displacement $\delta{Q}$. This can be understood by the fully symmetric nature of the Raman tensor of the $A_{1g}$ phonon that is isotropic in the $ab$-plane \cite{Wu2020}. Thus, we conclude that the anisotropy in $\Delta{R}_\textnormal{CP}$ should stem from the optical responses to the $A_{1g}$ phonon ($\partial{R}/\partial{Q}$) \cite{Chauviere2011,Baum2018}.
 
 We suggest distinct responses from $d_{yz}$ and $d_{zx}$ orbitals to the $A_{1g}$ motions as a possible origin for the anisotropy in the ($\partial{R}/\partial{Q}$). According to the optical selection rule, our probe light polarization, parallel to the nearest neighboring Fe-Fe bonding directions $x$($y$) (Fig. 1(b)), induces electronic transitions between $d_{xz}$($d_{yz}$) and $e_g$, and between $d_{yz}$($d_{zx}$) and $d_{xy}$ states \cite{Liu2018}. Without a nematic order, equivalent population in $d_{yz}$ and $d_{zx}$ orbitals makes optical responses of the P1 and P2 probes identical. However, the response becomes distinct in the nematic ordered phase as the degeneracy between the $d_{yz}$ and $d_{zx}$ is lifted \cite{Yi2011}. In the nematic phase, therefore, we can expect that transient reflectivity oscillations vary with the probe polarization if $d_{yz}$ and $d_{zx}$ react differently to the 5.45 THz $A_{1g}$ motions of $\delta{Q}$, resulting in anisotropy of ($\partial{R}/\partial{Q}$).
 
 The $A_{1g}$ phonon of arsenic ions is a particular mode strongly coupled to the ground state of BaFe\textsubscript{2}As\textsubscript{2}. As shown in Fig. 1(b), the $A_{1g}$ phonon motions directly modulate the Fe-As-Fe bonding angle $\alpha$ that is a critical factor to determine the superconductivity, spin-density-wave, as well as electronic structures in pnitides \cite{Yndurain2011,Mizuguchi2010,Kim2012,Okazaki2018,Gerber2017}. Recently, time-resolved photoemission experiments revealed anti-phase oscillatory spectral signals between the electron and hole bands of BaFe\textsubscript{2}As\textsubscript{2} due to the $A_{1g}$ vibrations, revealing the strong coupling of electronic structures to the $A_{1g}$ phonon \cite{Okazaki2018}. The exact responses of $d_{yz}/d_{zx}$ orbitals to the $A_{1g}$ motions, however, were hard to distinguish due to weak modulations in their hole bands in the photoemission spectra \cite{Okazaki2018}, while our time-resolved reflectivity data suggest an anisotropic behavior in these $d_{yz}/d_{zx}$ orbitals. 
 
 Finally, we note the phase difference $\Delta_{phase}$ in BaFe\textsubscript{2}As\textsubscript{2} requires a great care in the discussion of coherent phonons observed in optical experiments. It has been so far believed that the generation mechanisms of coherent vibrations of the lattice coordinates directly determine the phase of relevant oscillations in optical probing signals \cite{Zeiger1992,Stevens2002,Riffe2007}. In an opaque material, the abrupt modification of the charge distribution due to the photon absorption can trigger displacive motions of ions, resulting in cosine-type phonon oscillations \cite{Zeiger1992}. Alternatively, in a transparent material, impulsive stimulated Raman scattering process can generate sine-type oscillations of atomic motions \cite{Dhar1994,Stevens2002}. In general, the contribution of both mechanisms may result in an arbitrary oscillation phase. However, recent ultrafast spectroscopic studies showed huge variations in the phase of coherent phonon signals in optical properties depending on the electronic and magnetic ground states of correlated materials \cite{Lee2019_1,Lee2019_2}. These phase variations cannot be fully understood by the above mentioned generation mechanisms of coherent phonons, as the optical properties and phonon dephasing time in these complex materials did not manifest apparent changes across the phase transitions. Furthermore, the continuous variation of $\Delta_{phase}$ of BaFe\textsubscript{2}As\textsubscript{2}, i.e, the observation of different oscillation phases in two polarization directions although the oscillations originate from the same atomic vibrations of the $A_{1g}$ mode indicates that there exists another factor to determine the phase of coherent phonon oscillations in optical signals. These results may be attributed to retardation of electronic and/or optical responses to lattice vibrations on a femtosecond timescale, but it requires further investigation. 

 In summary, we investigate coherent phonon oscillations in photoinduced reflectivity of BaFe\textsubscript{2}As\textsubscript{2}, showing a clear anisotropic behavior related to the in-plane orthorhombic crystal axes. We reveal that the anisotropy in the oscillation-phase of the $A_{1g}$ coherent phonon clearly remains in the tetragonal paramagnetic state, which may originate from the nematicity with \textit{d}\textsubscript{yz/zx} orbital fluctuations. Our results suggest that coherent phonon oscillations can provide a critical evidence of a nematic order, and its coupling to the lattice structure in strongly correlated materials such as BaFe\textsubscript{2}As\textsubscript{2}.
 
 This work was supported by the Institute for Basic Science (IBS) in Korea (IBS-R009-D1). K.W.K. was supported by the National Research Foundation of Korea (NRF) grant funded by the Korea government (MSIT) (No. 2020R1A2C3013454 and No. 2020R1A4A1019566).


\begin{references}
\bibitem{Paglione2010} J. Paglione and R. L. Greene, \href{http://doi.org/10.1038/nphys1759}{Nat. Phys. \textbf{6}, 645 (2010).}
\bibitem{Johnston2010} D. C. Johnston, \href{https://doi.org/10.1080/00018732.2010.513480}{Adv. Phys. \textbf{59}, 803 (2010).}
\bibitem{Fernandes2014}	R. M. Fernandes, A. V. Chubukov and J. Schmalian, \href{http://doi.org/10.1038/nphys2877}{Nat. Phys. \textbf{10}, 97 (2014).}
\bibitem{Fernandes2012}	R. M. Fernandes, A. V. Chubukov, J. Knolle, I. Eremin, and J. Schmalian, \href{https://doi.org/10.1103/PhysRevB.85.024534}{Phys. Rev. B \textbf{85}, 024534 (2012).}
\bibitem{Chu2008} J.-H. Chu, J. G. Analytis, K. D. Greve, P. L. McMahon, Z. Islam, Y. Yamamoto, I. R. Fisher, \href{http://doi.org/10.1126/science.1190482}{Science \textbf{329}, 824 (2008).}
\bibitem{Kasahara2012} S. Kasahara, H. J. Shi, K. Hashimoto, S. Tonegawa, Y. Mizukami, T. Shibauchi, K. Sugimoto, T. Fukuda, T. Terashima, A. H. Nevidomskyy, and Y. Matsuda, \href{https://doi.org/10.1038/nature11178}{Nature \textbf{486}, 382 (2012).}
\bibitem{Yi2011} M. Yi \textit{et al}., \href{https://doi.org/10.1073/pnas.1015572108}{Proc. Natl. Acad. Soc. USA \textbf{108}, 6878 (2011).}
\bibitem{Kim2013} Y. K. Kim, W. S. Jung, G. R. Han, K.-Y. Choi, C.-C. Chen, T. P. Devereaux, A. Chainani, J. Miyawaki, Y. Takata, Y. Tanaka, M. Oura, S. Shin, A. P. Singh, H. G. Lee, J.-Y. Kim, and C. Kim, \href{https://doi.org/10.1103/PhysRevLett.111.217001}{Phys. Rev. Lett. \textbf{111}, 217001 (2013).}
\bibitem{Nakajima2011}  Nakajima, T. Liang, S. Ishida, Y. Tomioka, K. Kihou, C. H. Lee, A. Iyo, H. Eisaki, T. Kakeshita, T. Ito, and S. Uchida, \href{https://doi.org/10.1073/pnas.1100102108}{Proc. Natl. Acad. Soc. USA \textbf{108}, 12238 (2011).}
\bibitem{Mirri2015} C. Mirri, A. Dusza, S. Bastelberger, M. Chinotti, L. Degiorgi, J.-H. Chu, H.-H. Kuo, and I. R. Fisher, \href{https://doi.org/10.1103/PhysRevLett.115.107001}{Phys. Rev. Lett. \textbf{115}, 107001 (2015).}
\bibitem{Zeiger1992} H. J. Zeiger, J. Vidal, T. K. Cheng, E. P. Ippen, G. Dresselhaus, and M. S. Dresselhaus, \href{https://doi.org/10.1103/PhysRevB.45.768}{Phys. Rev. B \textbf{45}, 768 (1992).}
\bibitem{Kuznetsov1994} A. V. Kuznetsov and C. J. Stanton, \href{https://doi.org/10.1103/PhysRevLett.73.3243}{Phys. Rev. Lett. \textbf{73}, 3243 (1994).}
\bibitem{Lee2021} M.-C. Lee \textit{et al}., manuscript under preparation.
\bibitem{Stevens2002} T. E. Stevens, J. Kuhl, and R. Merlin, \href{https://doi.org/10.1103/PhysRevB.65.144304}{Phys. Rev. B \textbf{65}, 144304 (2002).}
\bibitem{Riffe2007} D. M. Riffe and A. J. Sabbah, \href{https://doi.org/10.1103/PhysRevB.76.085207}{Phys. Rev. B \textbf{76}, 085207 (2007).}
\bibitem{Lee2019_1} M.-C. Lee, C. H. Kim, I. Kwak, C. W. Seo, C. H. Sohn, F. Nakamura, C. Sow, Y. Maeno, E.-A. Kim, T. W. Noh, and K. W. Kim., \href{https://doi.org/10.1103/PhysRevB.99.144306}{Phys. Rev. B \textbf{99}, 144306 (2019).}
\bibitem{Lee2019_2} M.-C Lee, I. Kwak, C. H. Kim, B. Lee, B. C. Park, J. Kwon, W. Kyung, C. Kim, T. W. Noh, and K. W. Kim, \href{https://doi.org/10.1103/PhysRevB.100.235139}{Phys. Rev. B \textbf{100}, 235139 (2019).}	
\bibitem{Gerber2015} S. Gerber, K. W. Kim, Y. Zhang, D. Zhu, N. Plonka, M. Yi, G. L. Dakovski, D. Leuenberger, P. S. Kirchmann, R. G. Moore, M. Chollet, J. M. Glownia, Y. Feng, J.-S. Lee, A. Mehta, A. F. Kemper, T. Wolf, Y.-D. Chuang, Z. Hussain, C.-C. Kao, B. Moritz, Z.-X. Shen, T. P. Devereaux and W.-S. Lee , \href{http://doi.org/10.1038/ncomms8377}{Nat. Commun. \textbf{6}, 7377 (2015).}
\bibitem{Rettig2015} L. Rettig, S. O. Mariager, A. Ferrer, S. Gr\"ubel, J. A. Johnson, J. Rittmann, T. Wolf, S. L. Johnson, G. Ingold, P. Beaud, and U. Staub, \href{https://doi.org/10.1103/PhysRevLett.114.067402}{Phys. Rev. Lett. \textbf{114}, 067402 (2015).}
\bibitem{Yang2014} L. X. Yang, G. Rohde, T. Rohwer, A. Stange, K. Hanff, C. Sohrt, L. Rettig, R. Cort\'es, F. Chen, D. L. Feng, T. Wolf, B. Kamble, I. Eremin, T. Popmintchev, M. M. Murnane, H. C. Kapteyn, L. Kipp, J. Fink, M. Bauer, U. Bovensiepen, and K. Rossnagel, \href{https://doi.org/10.1103/PhysRevLett.112.207001}{Phys. Rev. Lett. \textbf{112}, 207001 (2014).}
\bibitem{Okazaki2018} K. Okazaki, H. Suzuki, T. Suzuki, T. Yamamoto, T. Someya, Y. Ogawa, M. Okada, M. Fujisawa, T. Kanai, N. Ishii, J. Itatani, M. Nakajima, H. Eisaki, A. Fujimori, and S. Shin, \href{https://doi.org/10.1103/PhysRevB.97.121107}{Phys. Rev. B \textbf{97}, 121107(R) (2018).}
\bibitem{Mansart2010} B. Mansart, D. Boschetto, A. Savoia, F. Rullier-Albenque, F. Bouquet, E. Papalazarou, A. Forget, D. Colson, A. Rousse, and M. Marsi, \href{https://doi.org/10.1103/PhysRevB.82.024513}{Phys. Rev. B \textbf{82}, 024513 (2010).}
\bibitem{Kim2012} K. W. Kim, A. Pashkin, H. Sch{\"a}fer, M. Beyer, M. Porer, T. Wolf, C. Bernhard, J. Demsar, R. Huber and A. Litenstorfer, \href{http://doi.org/10.1038/nmat3294}{Nat. Mater. \textbf{11}, 497 (2012).}
\bibitem{Stojchevska2012} L. Stojchevska, T. Mertelj, J.-H. Chu, I. R. Fisher, and D. Mihailovic, \href{https://doi.org/10.1103/PhysRevB.86.024519}{Phys. Rev. B \textbf{86}, 024519 (2012).}
\bibitem{Luo2017} C.-W. Luo, P. C. Cheng, S.-H. Wang, J.-C. Chiang, J.-Y. Lin, K.-H. Wu, J.-Y. Juang, D. A. Chareev, O. S. Volkova and A. N. Vasiliev, \href{https://doi.org/10.1038/s41535-017-0036-5}{npj Quant. Mater. \textbf{2}, 32 (2017).}
\bibitem{Liu2018} S. Liu, C. Zhang, Q. Deng, H.-H. Wen, J.-X. Li, E. E. M. Chia, X. Wang, and M. Xiao, \href{https://doi.org/10.1103/PhysRevB.97.020505}{Phys. Rev. B \textbf{97}, 020505(R) (2018).}
\bibitem{Rahlenbeck2009} M. Rahlenbeck, G. L. Sun, D. L. Sun, C. T. Lin, B. Keimer, and C. Ulrich, \href{https://doi.org/10.1103/PhysRevB.80.064509}{Phys. Rev. B \textbf{80}, 064509 (2009).}
\bibitem{Litvinchuk2008} A. P. Litvinchuk, V. G. Hadjiev, M. N. Iliev, B. Lv, A. M. Guloy,
and C. W. Chu, \href{https://doi.org/10.1103/PhysRevB.78.060503}{Phys. Rev. B \textbf{78}, 060503(R) (2008).}
\bibitem{Wu2020} S.-F. Wu, W.-L. Zhang, V. K. Thorsmolle, G. F. Chen, G. T. Tan, P. C. Dai, Y. G. Shi, C. Q. Jin, T. Shibauchi, S. Kasahara, Y. Matsuda, A. S. Sefat, H. Ding, P. Richard, and G. Blumberg, \href{ttps://doi.org/10.1103/PhysRevResearch.2.033140}{Phys. Rev. Research \textbf{2}, 033140 (2020).}
\bibitem{Chauviere2011} L. Chauvi{\'e}re, Y. Gallais, M. Cazayous, M. A. M{\'e}asson, A. Sacuto, D. Colson and A. Forget, \href{http://doi.org/10.1103/PhysRevB.84.104508}{Phys. Rev. B \textbf{84}, 104508 (2011).}
\bibitem{Baum2018} A. Baum, Ying Li, M. Tomi{\'c}, N. Lazarevi{\'c}, D. Jost, F. L{\"o}ffler, B. Muschler, T. B{\"o}hm, J.-H. Chu, I. R. Fisher, R. Valent{\'i}, I. I. Mazin, and R. Hackl, \href{http://doi.org/10.1103/PhysRevB.98.075113}{Phys. Rev. B \textbf{98}, 075113 (2018).}
\bibitem{Yndurain2011} F. Yndurain, \href{http://doi.org/10.1209/0295-5075/94/37001}{Euro. Phys. Lett. \textbf{94}, 37001 (2011).}
\bibitem{Mizuguchi2010} Y. Mizuguchi, Y. Hara, K. Deguchi, S. Tsuda, T. Yamaguchi, K. Takeda, H. Kotegawa, H. Tou, and Y. Takano, \href{http://doi.org/10.1088/0953-2048/23/5/054013}{Supercond. Sci. Technol. \textbf{23}, 054013 (2010).}
\bibitem{Gerber2017} S. Gerber, S.-L. Yang, D. Zhu,H. Soifer, J. A. Sobota, S. Rebec, J. J. Lee, T. Jia, B. Moritz, C. Jia, A. Gauthier, Y. Li, D. Leuenberger, Y. Zhang, L. Chaix, W. Li, H. Jang, J.-S. Lee, M. Yi, G. L. Dakovski, S. Song, J. M. Glownia, S. Nelson, K. W. Kim, Y.-D. Chuang, Z. Hussain, R. G. Moore, T. P. Devereaux, W.-S. Lee, P. S. Kirchmann, Z.-X. Shen, \href{http://doi.org/0.1126/science.aak9946}{Science \textbf{357}, 71 (2017).}
\bibitem{Dhar1994}  L. Dhar, J. A. Rogers, and K. A. Nelson, \href{http://doi.org/10.1021/cr00025a006}{Chem. Rev. \textbf{94}, 157 (1994).}
\end{references}
\end{document}